\documentclass[journal]{IEEEtran}

\usepackage{lipsum}
\usepackage[hidelinks]{hyperref}
\usepackage[binary-units=true]{siunitx}  
\usepackage{graphicx}
\usepackage{amsmath}
\usepackage{booktabs}
\usepackage{cite}
\usepackage{amssymb}
\usepackage{multirow}
\usepackage[normalem]{ulem}
\usepackage{textcomp}
\usepackage{fp}
\usepackage{pgfplots}
\pgfplotsset{compat=1.13}
\usepackage{pgfplotstable}
\usepackage{capt-of}
\definecolor{cieee0}{HTML}{00629b}
\definecolor{cieee1}{RGB}{255,199,44}
\definecolor{cieee2}{RGB}{232,119,34}
\definecolor{cieee3}{RGB}{186,12,47}
\definecolor{cieee4}{RGB}{119, 37, 131}
\definecolor{cieee5}{RGB}{120, 190, 32}
\definecolor{cieee6}{RGB}{0, 132, 61}
\definecolor{cieee7}{RGB}{0,  159,  223}
\usepackage{hyperref}

\newcommand{\secref}[1]{Section~\ref{#1}}
\newcommand{\tblref}[1]{Table~\ref{#1}}
\newcommand{\figref}[1]{Figure~\ref{#1}}

\graphicspath{{./figs/}}

\begin{document}

\title{\fontsize{23pt}{23pt}\selectfont Sub-mW Keyword Spotting on an MCU: Analog Binary Feature Extraction and Binary Neural Networks}

\author{
Gianmarco Cerutti, Lukas Cavigelli, Renzo Andri, Michele Magno, Elisabetta Farella, Luca Benini
\thanks{
    Gianmarco Cerutti and Elisabetta Farella are with ICT-irst Fondazione Bruno Kessler, Trento, Italy. (email: \{gcerutti, efarella\}@fbk.eu).
    
    Lukas Cavigelli and Renzo Andri are with Huawei Technologies, Computing Systems Laboratory, Zurich Research Center, Switzerland. (email: \{lukas.cavigelli, renzo.andri\}@huawei.com)
    
    Michele Magno and Luca Benini are with ETH~Zurich, Switzerland.
    (email: \{magnom, benini\}@ethz.ch)
    Luca Benini is also with the University of Bologna, Italy. (email:  luca.benini@unibo.it)
    }
}

\maketitle

\begin{abstract}
Keyword spotting (KWS) is a crucial function enabling the interaction with the many ubiquitous smart devices in our surroundings, either activating them through wake-word or directly as a human-computer interface. For many applications, KWS is the entry point for our interactions with the device and, thus, an always-on workload. Many smart devices are mobile and their battery lifetime is heavily impacted by continuously running services. KWS and similar always-on services are thus the focus when optimizing the overall power consumption. 

This work addresses KWS energy-efficiency on low-cost microcontroller units (MCUs). We combine analog binary feature extraction with binary neural networks. By replacing the digital preprocessing with the proposed analog front-end, we show that the energy required for data acquisition and preprocessing can be reduced by 29$\times$, cutting its share from a dominating 85\% to a mere 16\% of the overall energy consumption for our reference KWS application.

Experimental evaluations on the Speech Commands Dataset show that the proposed system outperforms state-of-the-art accuracy and energy efficiency, respectively, by 1\% and 4.3$\times$ on a 10-class dataset while providing a compelling accuracy-energy trade-off including a 2\% accuracy drop for a 71$\times$ energy reduction.
\end{abstract}

\section{Introduction}

Keyword spotting has become increasingly popular over the last few years with the wide-spread adoption in commercial products such as the “Ok, Google” and “Hey, Siri” commands used to wake up a smartphone and following up with a more complicated voice command \cite{michaely2017keyword}, the call for “Alexa” to activate Amazon's smart home devices, as well as many different keywords to activate recent cars' driver assistance systems. A key property of all these applications is their always-on nature and the high accuracy requirements, which are essential for the user experience. As voice command interfaces penetrate extremely cost-sensitive markets, such as toys and home-automation, keyword spotting is ported to low-cost MCU devices, with tight constraints in memory and computational power.

State-of-the-art keyword spotting methods rely on deep neural networks (DNNs) as an essential part of the processing pipeline. DNNs are known for their high compute effort and often have memory requirements in the megabyte to gigabyte range, making it a natural choice to off-load their compute workload to cloud services in data centers rather than running on power- and memory-constrained devices \cite{mohammadi2018deep}. This implies continuously sending a stream of audio data to a remote system and thus comes with many drawbacks: privacy concerns \cite{takabi2010security}, short battery lifetimes due to energy cost of data transmission \cite{alioto2017iot}, communication infrastructure and cloud computing service cost, availability in areas of poor connectivity, and a high latency \cite{dillon2010cloud}. 

These obstacles can be overcome by processing the data directly on the device where the sensor data is collected, avoiding transmitting the data entirely, or sending out only the essential information as part of an alert \cite{yogi2017mist, premsankar2018edge, krestinskaya2018binary}. This requires a carefully designed system, combining the latest methods on constructing efficient and compact DNN models considering constraints on the compute effort and memory requirements with the latest DNN deployment methods for efficient inference to keep the energy spent on computing minimal for long battery life. 

Many methods have been proposed in pursuit of efficient DNN inference, when targeting MCU devices, from neural architecture search to knowledge distillation and novel efficient building blocks for DNNs. A particularly successful method to bring keyword spotting to embedded devices has been the training of extremely quantized DNNs known as binary neural networks (BNNs), where the large compute workload performing multiply-add operations in convolution layers decays to XNOR and popcount operations, and a data word in memory can hold a vector of weight or feature map values. This has shown success in many applications, including keyword spotting \cite{cerutti2020sound}.
Multiply-accumulate operations are dominating the compute effort of neural networks, thus measuring the throughput of a system in terms of GMAC/s and its energy efficiency with GMAC/s/W has been widely adopted as a network-agnostic performance metric. Nevertheless, when quantizing neural networks to few-bit operands, MAC operations become significantly simpler, and larger networks are required to compensate the accuracy drop. The energy efficiency in the form of energy per classification and throughput measured as classifications per seconds are considered more representative metrics for real world performance assessment \cite{burr2021fair}.

A complete keyword spotting system consists of many components. With a strongly optimized classifier, classification computation energy drops from being the dominant contributor to third place after data acquisition (microphone, analog filtering, analog-to-digital conversion) and pre-processing (computation of Mel spectrograms). Audio classification systems are often based on the regular sampling of microphone inputs with frequencies ranging from 16 kHz to 192 kHz in order to preserve the spectral information from the audio stream as much as possible~\cite{panayotov2015librispeech, mesaros2018multi}. The sampling rate dictates a lower bound on continuous power consumption.
However, successful classification requires only suitable features and not necessarily the entire raw data stream. The overall energy efficiency can thus be improved by a sensor or acquisition block that returns only the relevant features for classification.

In this paper, we combine a binary neural network for efficient classification on MCUs with an optimized feature extractor in the analog domain leading directly to \emph{binary} time-frequency features, eliminating the need for digital pre-processing and analog-to-digital conversion. We use an active filter bank followed by envelope detectors and comparator circuits to indicate activity in each of the frequency bands. The resulting digital signals are then used as interrupts for the MCU, signaling any change of the active frequency bands. This allows putting the MCU in sleep mode most of the time, further lowering the continuous power usage by following the event-based acquisition paradigm. 
Our contributions can be summarized as follows:
\begin{itemize}
    \item We combine a power-optimized analog front end with direct binary feature extraction suitable for keyword spotting.
    \item We implement and characterize all components of the keyword spotting system, including: analog front-end with interrupt-based wake-up, data acquisition, pre-processing, and optimized BNN inference, implemented on a low-power multi-core RISC-V-based MCU. 
    \item We explore the impact of various degrees of freedom in our analog front end, specifically the filter bank size and filter properties, and the resulting impact on the final classification accuracy. 
    \item We thoroughly analyze the trade-offs between accuracy and energy efficiency, including sweeps of the filter bank properties as well as the BNN size.
    \item We compare the proposed system in-depth with state-of-the-art methods, showing how it can be parametrized to span the entire energy-accuracy Pareto front. 
\end{itemize}

\section{Related Work}
Deep neural networks are the state-of-the-art technique for audio and speech processing. Their specific type varies from convolutional neural networks (CNNs) for extracting keywords and phonemes from spectrogram-like inputs \cite{Sainat2015cnnkws} to recurrent neural networks (RNNs) such as LSTMs and GRUs \cite{lengerich2016end, hwang2015online}, spiking neural networks (SNNs) \cite{pedroni2018small}, temporal convolution networks (TCNs) \cite{Pandey2019}, and ultimately attention-based models with large embeddings \cite{Vaswani2017} as the tasks move from analyzing sounds to understanding entire sentences with wide contexts.

\subsection{Size-Optimized \& Quantized DNNs}

Keyword spotting systems are typically mobile edge devices, where the data is processed on the device to keep the energy cost acceptable and overcome the privacy concerns of the users that would come with the transmission of their data to the cloud. 
A lot of research has thus been conducted in the direction of optimizing the DNN models to fit the device constraints, either by manually exploring various DNN models specifically for keyword spotting, or through more general research on how to find compute-efficient DNNs. 

A typical approach to coping with resource constraints is to reduce the model size to fit these devices. For example, the authors of \cite{Sainat2015cnnkws} test different architectures to explore the consequence of changing the size of different layers. On the same path, Tucker et al. \cite{tucker2016model} investigated the use of low-rank weight matrices and knowledge distillation. However, they try to increase the performance without changing the model size.

An additional level of optimization relies on changing how numbers are represented. Deep learning frameworks train neural networks using 32-bit floating-point weights and activations. Researchers have tried to reduce numerical precision to save memory, reduce internal data bandwidth, and simplify the compute operations. This process, called quantization, aims to reduce the storage and computational costs of the inference task \cite{lin2016fixed}. On the other hand, the performance of the network in terms of accuracy can degrade. If properly trained, the weight-quantized networks can achieve an accuracy close to the floating-point original models, also on complex classification tasks \cite{Leng_2018, liu202022nm}. Zhou et al. \cite{AojunZhou2016} present a technique for lossless weight quantization down to 2 bits.  This approach saves memory and bandwidth, but keeping activations in full-precision always requires floating-point computations and 32 bit for each activation.
Unfortunately, activations are more sensitive to quantization in terms of accuracy drop. The approximation of the activations is quasi-lossless down to 8 bit \cite{zhang2017hello}, and it allows the hardware to parallelize four computations over a 32-bit register. 

As an extreme case of quantization, binary neural networks (BNNs) reduce the precision of both weights and neuron activations to a single-bit \cite{Courbariaux2016, rastegari2016xnor}. BNNs work well on simple tasks like MNIST, CIFAR-10, and SVHN without impacting the accuracy \cite{hubara2016binarized}, but are showing worse performance on challenging datasets such as ImageNet with a drop of around 12\% \cite{zhou2016dorefa, spallanzani2019additive}. BNNs provide major benefits for computation. Binarization reduces the amount of memory required for both the weights and the intermediate results, fitting 32 values into a single 32\,bit word. Further, such a vectorized representation in memory can be used directly to compute 32 MAC operations with just two light-weight instructions---the multiplication simplifies to a bit-wise XNOR operation, and the summing of the 32 values can be done with a popcount instruction.

\subsection{KWS Applications \& Implementations}

Focusing on always-on keyword spotting, the authors of \cite{gruenstein2017cascade} reduce the power consumption with multiple stages that wake up a more complex system. The first stage comprises a tiny and power-efficient detector that executes on a DSP. Upon trigger, it delegates the final detection decision to a second, much larger, and more accurate detector.

Keyword spotting is not limited to high-end consumer devices: many interesting applications belong to the IoT domain, where the end nodes are typically MCU-based sensor devices with even more limited resources than a mobile phone. A wireless sensor node needs to have a compact form factor and very low cost. This requirement also limits the battery's size, thus setting the bound for the power consumption to a few milliwatts. 

Pioneers of edge computing for keyword spotting \cite{zhang2017hello} manage to implement a keyword spotting model on a cortex M4 based MCU: they explore a different kind of models, and they reach the state of the art accuracy even in memory-constrained devices. Moreover, they present an implementation with 8-bit quantization for weights and activations to use SIMD instruction and parallelize four operations in just one instruction.
Following this trend, Justice et al. \cite{amoh2019optimized} proposed a 3-bit quantization for the weights while keeping 16 bits for the activations. They also present a recurrent unit for keyword spotting, demonstrating its effectiveness in execution time and memory footprint. Here, they use a Cortex M0 MCU with a memory size of 32\,kB. 
The authors of \cite{cerutti2020sound} build on top of \cite{meyer2017efficient} by going down to 1-bit weights and activations for audio event detection, and they demonstrate that thanks to BNN, the system achieves an efficiency of 31.3\,GMAC/s/W thanks to parallelization over eight cores and an optimized Instruction Set Architecture.

All these implementations achieve acceptable accuracy and small memory footprint for their models, but they do not focus on the overall system's power consumption. In fact, the microphone itself, its amplification circuitry, and the communication between the MCU and the sensor are a relevant part of the energy used for keyword spotting or audio event detection. Cerutti et al. \cite{cerutti2020compact} present an edge-computing system for sound event detection, and they conclude that the system uses more than 60\% of the power budget for sensor reading and feature extraction, and only the remaining part is for classification.

To improve data acquisition efficiency, a vast amount of research prototypes try to mimic human perception, creating a time-frequency representation of the microphone output, an approach known as silicon cochlea.
\cite{wen2006360, fragniere2005100, lyon1988analog, 6869048, 5537160,li2021ns}. Promising features in this approach are event-based processing and avoidance of traditional Nyquist sampling. 
In \cite{5537160,yang20181muw}, the authors design a custom chip for voice activity detection which integrates both the feature extractor as well as the neural network. The audio signal passes through a bank of hardware filters, and sequentially, an integrator generates a set of spikes according to the power present in the specific frequency band. Voice activity detection is generally more manageable than a multi-class classification like keyword spotting. Thus, they can get a reasonable accuracy with only a 3-layer neural network  with 48 inputs and 4.6\,kB of parameters.

\subsection{Integrated KWS SoCs}
\label{sec:relWork_integrated}
We focus primarily on a full system implementation based on commercially available components. However, fully-integrated implementations can massively reduce the power requirements and some innovative, state-of-the-art KWS system concepts are introduced the following works. 

With Vocell \cite{giraldo2020vocell}, Giraldo et al. propose a system for KWS and speaker verification, integrating the AFE and a cascade of a sound detector, feature extractor, KWS classifier, and speaker verification units into a single SoC. Its average measured power consumption varies from 6.5\,\si{\micro\watt} (90\% silence) to 18.3\,\si{\micro\watt} (continuously running), uses 16\,\si{\micro\watt} on average for 500\,ms in KWS-mode with extreme voltage scaling (0.6\,V for AFE, 0.3\,V for logic), and it achieves 90.87\% accuracy for KWS on the Google Speech Commands dataset with 12 classes. 

In the most recent work of Giraldo et al., they present a temporal convolutional network (TCN) accelerator achieving close-to-SoA performance of 93.3\% on the GSCD task requiring 1.5\,INT8-MOp at 8\textmu W power consumption. \cite{giraldo2021efficient}. Shan et al. \cite{shan202014} show the smallest power numbers, proposing an accelerator for MFCC, binarized depthwise-separable convolutions consuming 510 nW, but on a very simple dataset (1 or 2 words of GSCD). Chong et al. present another MFCC and LSTM accelerator, by exploiting weight-stationary paradigm, small FFT sizes, rectangular Mel filters, and highly-pruned LSTM model (89\%), they achieve 2.5\textmu W at 400\,kHz and 0.6\,V supply (excluding microphone, ADC, I/O power) at 90.6\% accuracy on a 10-class GSCD \cite{chong20212}.

H. Fuketa \cite{fuketa2021time} addresses the same issue, the energy cost of ADCs, and proposes to move the pre-processing to the analog domain. They replace the ADC+MFCC front-end with a filter bank of 10 band-pass switched-capacitor filters, followed by an analog maximum sampling circuit (100\,Hz), a tiny convolutional neural network layer to convert 2 samples (20 values) to a 64-value vector at 50\,Hz that is then binarized by comparators and finally processed by binary time-delay neural network (TDNN) in digital domain. They achieve an 88.8\% accuracy when trained on a subset of only 3 classes (`yes', `no', silence) of the Google Speech Commands dataset, while using merely 0.25\,\si{\micro\watt} for the entire simulated circuit (from AFE to classification result).

M. Yang et al. \cite{9429864} also propose to move pre-processing to the analog domain where they pass the signal through a low-noise amplifier, a 16-channel bandpass filter followed by clipping amplifier, a half-wave rectifier, and an integrate-and-fire encoder for ADC. Not aiming at high linearity in the AFE allows them to further reduce the power consumption to 53\,nW.

All the above works would require extensive qualification work and design margins to become mature for high-yield mass production, and in addition they do not consider the extra overhead linked to system integration (chip I/O, external power supplies, \dots). In this paper, we follow a full-system approach based on commercial off-the-shelf components already in mass production. We combine an ultra-low-power mixed-signal feature extractor front-end derived from \cite{mayer2019self}, with an optimized binary neural network running on a parallel low-power MCU. The front-end directly produces binary patterns used to wake up the MCU and are then fed directly as input to the binary neural network. This streamlined and energy-efficient interface between feature extraction and classification dramatically reduces power consumption while at the same time achieving compelling accuracy compared to SoA, as shown in \secref{sec:results}.

\begin{figure*}[tb]
  \includegraphics[width=\textwidth]{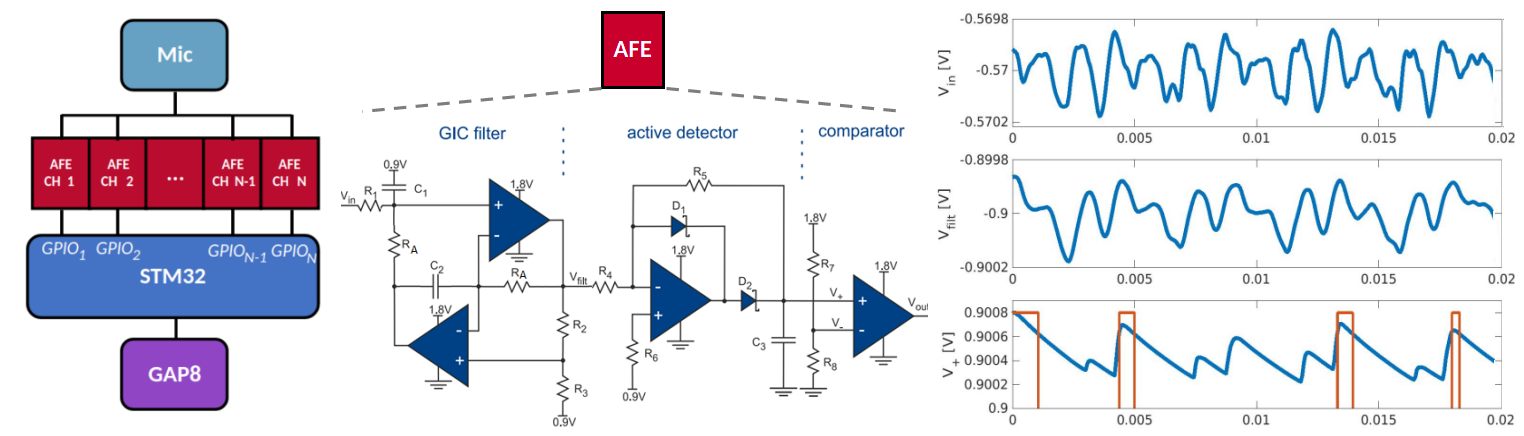}
  \caption{Left: overall structure of the keyword spotting system. Center: schematic of the analog front end, divided in its three main blocks. Right: intermediate signal in the circuit. From the top: the microphone output, the filtered signal and the output of the active detector. In red the binary output after the comparator.}
  \label{fig:AFE_schematic}
\end{figure*} 

\begin{figure}[tb]
    \centering
    \includegraphics[width=\columnwidth]{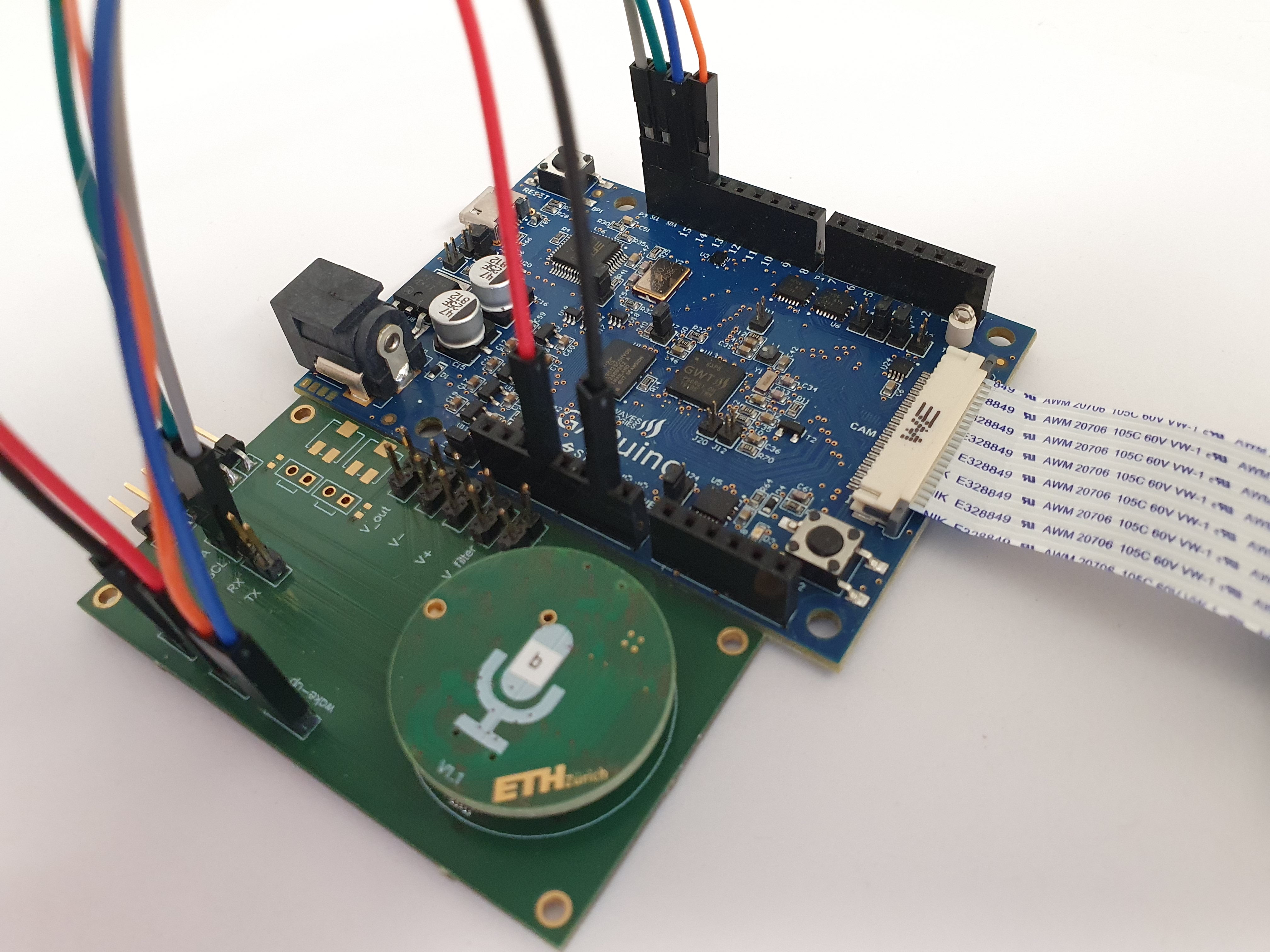}
    \caption{Picture of the system: acquisition board with the analog front end (bottom left) and the GAPuino processor board with the Greenwaves GAP-8 8+1 core RISC-V processor (top right). }
    \label{fig:system_picture}
\end{figure}

\section{Sensor and Binary Feature Sampling}
Three key components define a keyword spotting system: the sensing element, which converts the acoustic wave to an electrical signal; the feature extractors, which pre-process the data to obtain a more meaningful representation of the input; and a classifier, which performs classification using the previously generated features. In this paper, we exploit a low-power analog front-end to convert the microphone data directly to a binary signal, ready to be processed in a binary neural network. The following sections detail the subsystems.

\subsection{Analog Front-End}
In the proposed approach, data acquisition and the following feature extraction are performed using a low-power and energy-efficient analog front end, motivated by \cite{mayer2019self}. \figref{fig:AFE_schematic} shows the circuit schematic of the front end employed in this work. The analog signal processing allows the use of any analog microphone; we use a low-noise MEMS from InvenSense (ICS-40310), which consumes only \SI{16}{\micro\ampere} at \SI{0.9}{\volt}. The analog front end aims to extract frequency-time features, thus the next step is filtering the frequency bands of interest. 
The output of the microphone is connected to a general impedance converter (GIC), which uses the OPA379 operational amplifier from Texas Instruments. It features a gain-bandwidth product of \SI{90}{\kilo\hertz} and a current of \SI{2.9}{\micro\ampere} at \SI{1.8}{\volt}. The configuration generates a band-pass filter with a center frequency and corner frequency 
\begin{equation}
    f_c = \frac{1}{2\pi R_A C} \quad BW = f_c\frac{R_A}{R_1}.
\end{equation}
The analog front end has an active envelope detector to keep track of the temporal duration for which the frequency is active. It also amplifies the output of the GIC, following the formula
\begin{equation}
G = \frac{R_5}{R_4}.
\end{equation}
Finally, the envelope passes through a comparator. The LPV7215 consumes \SI{580}{\nano\ampere} at \SI{1.8}{\volt}. The first input is the output of the envelope detector, the second is a predefined reference voltage. 

\begin{figure}[tb]
  \includegraphics[width=0.95\columnwidth ]{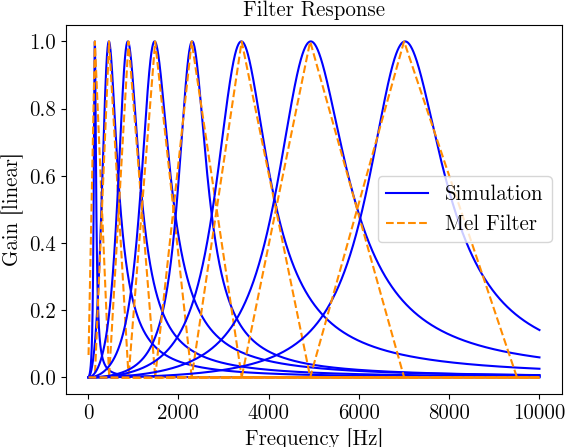}
  \caption{Frequency response (in blue) and Mel filter response when the analog front end has 8 filters.}\label{fig:freqresp}
\end{figure}

To achieve a high classification accuracy in a complex task such as the keyword spotting, more than a single time-frequency detector is required. Therefore the front-end need to be tuned with the resistor and capacitors in the GIC to select a different center frequency and bandwidth of the filter. In the proposed system, the same microphone is connected to all these filters.
The next step is to choose the right filter shape for further classification: log-Mel spectrogram and the following MFCC features are intensively used in sound-based classification systems \cite{zhang2017hello, Sainat2015cnnkws}. For this reason, we choose corner frequencies equally-spaced in the Mel domain within the selected range. We tuned the resistors as well to match the bandwidth of Mel filters.
\figref{fig:freqresp} shows the Mel filter responses and the filter banks' analog simulation. Even if the analog filters do not have a triangular shape, they are a close approximation.
\tblref{table:AFE_consumption} shows power consumption for different components and filter configurations for data acquisition and preprocessing with the analog front end with the conventional baseline approach. Overall, the data acquisition power can be reduced by 28$\times$.

\begin{table}[tb]
\centering
\caption{Analog Front End Power Consumption compared with conventional data acquisition and pre-processing}
\label{table:AFE_consumption}
\begin{tabular}{lrrr}
\toprule
System Component & \multicolumn{3}{r}{Energy} \\ \midrule
AFE with 64/16/8 filters       & 1248 & 306 & 153\,\si{\micro\joule} \\
\ - Microphone              & 288 & 72 & 36\,\si{\micro\joule} \\
\ - Single Time-Frequency Detector & 15 & 15 & 15\,\si{\micro\joule} \\ 
MCU consumption during acquisition & 291 & 168 & 131\,\si{\micro\joule} \\
Total Preprocessing (incl. everything) & 1539 & 488 & 284\,\si{\micro\joule}  \\
\midrule
baseline: MEMS microphone with ADC & \multicolumn{3}{r}{5400\,\si{\micro\joule}}  \\ 
baseline: SoA preprocessing (MFCC) & \multicolumn{3}{r}{2640\,\si{\micro\joule}} \\
\bottomrule
\end{tabular}
\end{table}

\subsection{Digital Sampling}\label{section:firmware}
The output of the analog front end contains two pieces of information for each filter. It shows the filter pass-band frequency component is present in the input data and how long it lasts. The binary output corresponds to the specific frequency band's presence; the length of the pulse provides the duration. For the following classification part, we want to save this information. 

The microcontroller has each digital output of the Analog Front End connected to a general purpose input-output (GPIO) pins. In a silent situation, the microcontroller is in sleep mode. When the first interrupt arrives from one of the channels, the microcontroller starts a timer. That will be considered time 0. From now on, the microcontroller saves the time stamps of each interrupt for all the channels. 

After one second, the firmware starts the reconstruction of the time-frequency representation from the collected data. 
As in a spectrogram, we discretize the time axis in windows. 
We fill the time-frequency representation with a ``1" if the analog front-end output is high at least once inside the window for each channel and each window. In this way, we directly generate the binary image that will be the input of the binary neural network. 

The power consumption in this phase is mainly given by the short interrupt service during which the microcontroller saves the timestamp of the event. To estimate this power, we run the firmware with a set of interrupt rate on real hardware and we estimate the power consumption per interrupt. Finally, we compute the number of interrupts in the simulations for each different setup (8, 16 and 64 filters). 
The platform used for this experiment is the STM32l476RG, and we configured the low-power timer to run with a 32 kHz clock.

For the baseline directly sampling the audio signal, we measured the power consumption of the data acquisition using an SPH0645 MEMS microphone and the SAI peripheral.
The acquisition of 1\,s of data requires 5.4\,mJ using the MEMS microphone, while with the analog front end with 8 filters the energy consumption is 0.28 mJ, i.e. just 5.2\% of the energy consumed by using the standard microphone.

\begin{figure}[ht]
  \includegraphics[width=0.95\columnwidth ]{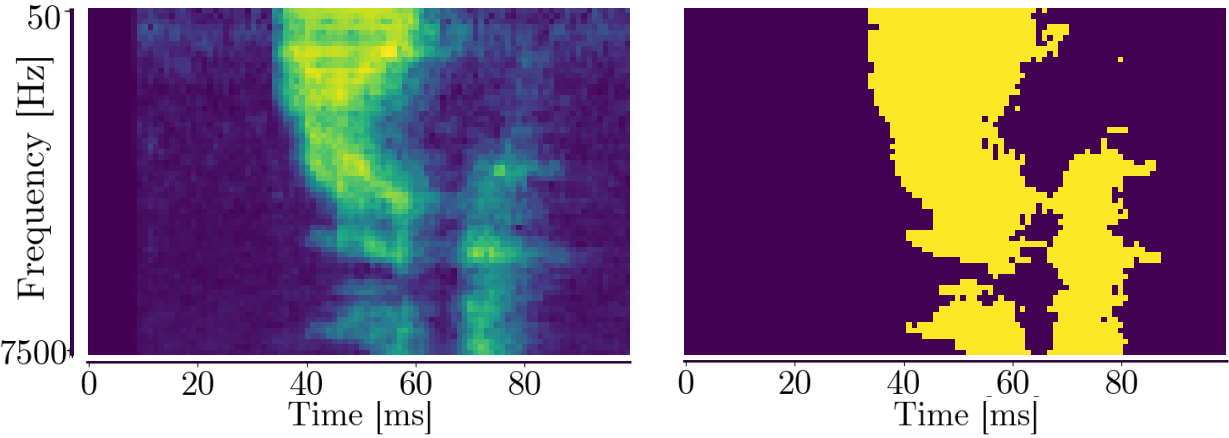}
  \caption{Simulation of the AFE with 64 filters (left) and its binarized version (right) of the keyword ``right''. The 64 filters are equally spaced in the logarithmic Mel domain between 50 and 7500 Hz.}\label{fig:bin_spectrogram}
\end{figure}

\section{Binary Neural Network}
\subsection{Dataset and Binarization}\label{sec:dataset}
To train and to evaluate the proposed system, we used the Google Speech Commands V2 dataset, which contains 105k speech samples of 35 words \cite{warden2018speech}. The neural network model is trained to classify the processed audio in one of the 10 keywords: ``Yes", ``No", ``Up", ``Down",``Left", ``Right", ``On", ``Off", ``Stop", ``Go". In addition, there is a class ``silence" (i.e., no word spoken) and ``unknown" word, which is a subset of the utterances in the remaining 20 keywords.
The train-validation-step split has an 80:10:10, and it is done following the recommendations of the dataset.

When considering the analog front end,  each clip of the dataset is converted in the AFE output. Meyer et al. demonstrated the similarity between the AFE output and a binarized spectrogram \cite{mayer2019self}.  This similarity motivates us to use filters to generate the binary features.
From these features, we obtain the maximum full-precision values of the spectrogram in windows of \SI{10}{\milli\second} or \SI{25}{\milli\second}. The binarization thresholds (one for each channel) are initialized with the single-channel average value. The full-precision envelopes' values are normalized using min-max scaling, and the same min-max values are used to scale the initial thresholds. The analog front end threshold can be adapted to this optimal threshold by selecting the corresponding resistor divider, which creates the desired voltage reference for the comparator. We visualize the intermediate presentations in \figref{fig:bin_spectrogram}, both before and after binarization, the latter being the input of the BNN. 

\subsection{Binary Convolution}\label{subsec:binaryconvolution}
Binary neural networks (BNNs) constrain both the weights and inputs to $\textbf{I}\in\{-1, 1\}^{n_{in}\times h\times b}$ and $\textbf{W} \in \{-1, 1\}^{n_{out}\times n_{in}\times k_y \times k_x} $. To avoid using two bits, we represent $-1$ with $0$, whereas the actual binary numbers are indicated with a hat (i.e., $\hat{i}=(i+1)/2$). It turns out that multiplications become \texttt{xnor} operations $\bar{\oplus}$ \cite{rastegari2016xnor}. Formally the output $o_k$ of an output channel $k\in \{0, ..., n_{out}-1\}$  can be described as\footnote{For simplicity, we omit the bias and the scaling factor in the formula.}:

\resizebox{0.87\linewidth}{!}{
  \begin{minipage}{\linewidth}
  \[
\mathbf{o_k} = \text{sgn}\left(\sum_{n=0}^{n_{in}-1}{{\mathbf{i_n} \ast \mathbf{w_{k,n}}}}\right) = \text{sgn}\left(\sum_{n=0}^{n_{in}-1}{2\left({\mathbf{\hat{i}_n} \ast \mathbf{\hat{w}_{k,n}}}\right)-k_yk_x}\right)\]
\[
= \text{sgn}\left(\sum_{n=0}^{n_{in}-1}{\sum_{(\Delta x, \Delta y)}{2\left({{\hat{i}_n}^{y\text{+}\Delta y,x\text{+}\Delta x} \bar{\oplus} {\hat{w}_{k,n}}}^{\Delta y, \Delta x}\right)-1}}\right)\]
  \end{minipage}
}
\newline\vspace{1mm}\newline
Whereas $\Delta y$ and $\Delta x$ are the relative filter tap positions (e.g., $(\Delta y, \Delta x)\in\{-1,0,1\}^2$ for $3\times 3$ filters). As calculating single-bit operations on the microcontroller is not efficient, we pack several input channels into a 32-bit integer (e.g., the feature map pixels at $(y+\Delta y,x+\Delta x)$ in spatial dimension and input channels $32n$ to $\left(32(n+1)-1\right)$ packed in $\mathbf{\hat{i}_{32n:+32}}^{y\text{+}\Delta y,x\text{+}\Delta x}$), while the Multiply Accumulates (MACs) can be implemented with \textit{popcount} and \textit{xnor} operations. 

Furthermore, as common embedded platforms like GAP8 do not have a built-in \textit{xnor} operator, the \textit{xor} operator $\oplus$ is used and the result is inverted. Therefore, the final equation for the output channel is $\mathbf{o_k}=$\newline 
\resizebox{1\linewidth}{!}{
  \begin{minipage}{\linewidth}
  \begin{align*}
\text{sgn}\left(\sum_{n=0}^{\frac{n_{in}}{32}-1}{\sum_{(\Delta x, \Delta y)}{32-2\text{popcnt}\left(\mathbf{\hat{i}_{32n:+32}}^{y\text{+}\Delta y,x\text{+}\Delta x} {\oplus} \mathbf{\hat{w}_{k,32n:+32}}^{\Delta y, \Delta x}\right)}}\right)
\end{align*}
  \end{minipage}
}
\newline\vspace{-4mm}\newline

\subsection{Batch Normalization and Binarization}
A batch normalization layer follows each binary convolutional layer. As the output of binary layers are integer values, and the signum function can be written as a comparison function, the activation function is simplified to:
\begin{equation}
\text{binAct}(x) = \begin{cases} 0, & \mbox{if } x \cdot \text{sgn}(\gamma') \geq \left\lfloor \frac{\beta'}{\gamma'} \right\rfloor \\[1mm]  1, & \mbox{if } x \cdot \text{sgn}(\gamma') < \left\lfloor \frac{\beta'}{\gamma'} \right\rfloor 
\end{cases}.
\end{equation}
whereas $\gamma'$ is the scaling factor, and $\beta'$ is the bias based on the batch normalization parameters. While exporting the model, we compute the integer threshold value $\lfloor\frac{\beta'}{\gamma'}\rfloor$ in advance.
In inference, one sign comparison and one threshold comparison have to be calculated for each activation value.

\subsection{Last Layer and Prediction}
In the last layer, the fixed-point values from the last binary layer are convolved with the fixed-point weights, and N output channels are calculated, where N is the number of classes. Finally, the network performs an average pooling over the whole image giving N predictions for each class.

\subsection{Network Architectures}
\label{sec:neural_network_architecture}
To investigate the trade-off in terms of accuracy and power consumption, we trained several networks with the same structure as \tblref{tbl:network_summary} but with a variable number of filters for each layer.
In particular, we have chosen a base vector, and then we have multiplied it by a scaling factor.   
To fully exploit the benefits of parallelization over 32 numbers, the minimum number of filters is 32. So we multiplied the base vector for 32, 64, and 128.
Different is the case of full-precision neural networks, where there is no reason to set a minimum number of filters. Therefore, we have multiplied the base vector by 4, 8, 16, and 32.
The last layer has to match the number of classes, so it does not change with the scaling factor.

\begin{table}[tb]
    \centering
    \caption{Neural Network Architecture: The base vector is multiplied by different values to change the channel count}
    \label{tbl:network_summary}
    \begin{tabular}{lccc}
    \toprule
    \textbf{Layer} & \textbf{Kernel Size} & \textbf{Stride} & \textbf{Base Vector} \\
    \midrule
    1. Conv Layer & 3 $\times$ 3& 2 &1\\
    2. Conv Layer & 3 $\times$ 3& 1 &2\\
    3. Conv Layer & 3 $\times$ 3& 2 &2\\
    4. Conv Layer & 3 $\times$ 3& 1 &3\\
    5. Conv Layer & 1 $\times$ 1& 1 &3\\
    6. Conv Layer & 1 $\times$ 1 & 1 &-\\
    \bottomrule
    \end{tabular}
\end{table}

\subsection{Microcontroller for BNN Inference}
For the estimation of the power consumption, we took the efficiency and throughput values from our previous work, which are 31.3\,GMAC/s/W and 1.5\,GMAC/s respectively \cite{cerutti2020sound}. The reference platform for BNN execution is the GAP8 platform. 
GAP8 is a commercial processor, implemented from the Parallel Ultra Low Power (PULP) open-source project\footnote{https://www.pulp-platform.org}. This processor has similar power requirements as the Cortex-M family with up to 20 times higher computation performance for machine learning applications, thanks to near-threshold parallel computing \cite{GapNews2018}. Furthermore, it features RISC-V ISA extensions for bit manipulation, effectively accelerating BNN inference with a native \texttt{popcount} instruction. 

\section{Results}\label{sec:results}
This paper comprises two novel contributions: the use of binary neural networks for keyword spotting and the combination with an analog front end that extracts time-frequency binary features. About the former, we first explore how the binarization of weights and activations impacts the classification performance.

\subsection{Effect of Model Binarization}
First, we trained a well-established convolutional neural network using standard log-Mel features as input to the network in \tblref{tbl:network_summary}. For the whole paper, it will be considered as a reference for other results. We use windows of \SI{25}{\milli\second} and a hop size of \SI{10}{\milli\second} to compute the spectrogram. We apply the Mel transformation using 64 Mel-filters in the range between 50 and 7500 Hz, following the methodology described in Hershey et al. \cite{hershey2017cnn}. The network is described in \tblref{tbl:network_summary} and the number of channels is obtained by multiplying the base vector by 64. Moreover, the layers are here in full-precision and not binary as depicted in the table. Between each convolution, there is Batch Normalization and ReLU activation. We used the Adam optimizer and a learning rate of $10^{-4}$ with a reduction of a factor of ten when the training is not improving for 10 consecutive epochs.
\begin{figure}[tb]
\centering
\pgfplotsset{width=10cm,compat=1.8, xmin=70, xmax=97}
\begin{tikzpicture}
 \begin{axis}[
  compat=newest,
  axis on top,
  xbar,
  bar width= 0.7cm,
  enlarge y limits=0.2,
  xmajorgrids,
  xlabel={Accuracy [\%]},
  y=1.0cm, xmin=60, xmax=100,
symbolic y coords={%
    {BNN binary input},
    {BNN},
    {Binary weights},
    Baseline},
  y tick label style={anchor=west,color=white,xshift= \pgfkeysvalueof{/pgfplots/major tick length}},
  bar shift=0pt,
  nodes near coords,axis on top=false, nodes near coords align={horizontal},
 ]
\addplot+[cieee4!20!black,fill=cieee4!60!white] coordinates {
    (85.6,{BNN binary input}) 
    (89.9,{BNN}) 
    (93.2,{Binary weights})
    (93.4,Baseline)};
 \node[text=white, anchor=west] at (axis cs: 60.5,{BNN binary input}) {BNN binary input};
 \node[text=white, anchor=west] at (axis cs: 60.5,{BNN}) {BNN};
 \node[text=white, anchor=west] at (axis cs: 60.5,{Binary weights}) {Binary weights};
 \node[text=white, anchor=west] at (axis cs: 60.5,{Baseline}) {Baseline};
 \end{axis}
\end{tikzpicture}\vspace{-4mm}
\caption{Effect of binarization of Mel-based classifier. The full-precision baseline is compared with versions trained with binary weights (BWN), binary weights and activations (BNN), and binary weights, activations and inputs (BNN with binary inputs) using thresholds learned during the training process.}\label{results:mel_bars}
\end{figure}
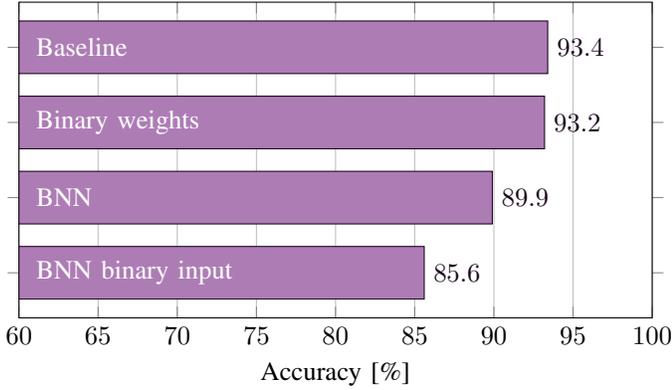

\figref{results:mel_bars} shows the performance impact of using binary numbers instead of full-precision ones. In detail, first, we binarize weights, activations, and input. In the first three cases, the first layer has full-precision input and weights, but the first activation function already binarizes the data. In the last case, with binary input, the Mel-spectrogram is compared with a channel-wise trainable threshold, and the first layer receives the binary spectrogram as input. The thresholds are initialized using the mean of each channel over the complete training set.
The second bar performs even better than the full-precision weight version; most likely, the binarization of weights acts as a regularization in this phase. Quantization of both weights and activations instead significantly decreases the accuracy. Finally, the approach of quantizing the input data using a threshold instead of a fully connected layer is the worst case, but it is the most similar case to the actual use of the analog front end. 

\subsection{Frequency Range Selection}
Here we present the analysis that we have carried out to choose the proper frequency range. From the recording setup, we know that the frequency limit is between 50 and 8000\,Hz. Having these constraints in mind, we looked for the best frequency range to select our corner frequencies for the filters. Once the range is fixed, we equally space the corner frequency in the Mel domain. We limit the analysis to the solution with eight filters.

Firstly, we used the full range of frequency; then, we explored which components are more relevant, excluding alternatively low and high frequency. Finally, we tried to exclude a smaller portion of the high and low spectrum. 

\begin{figure}[tb]
\centering
\includegraphics[width=0.98\columnwidth]{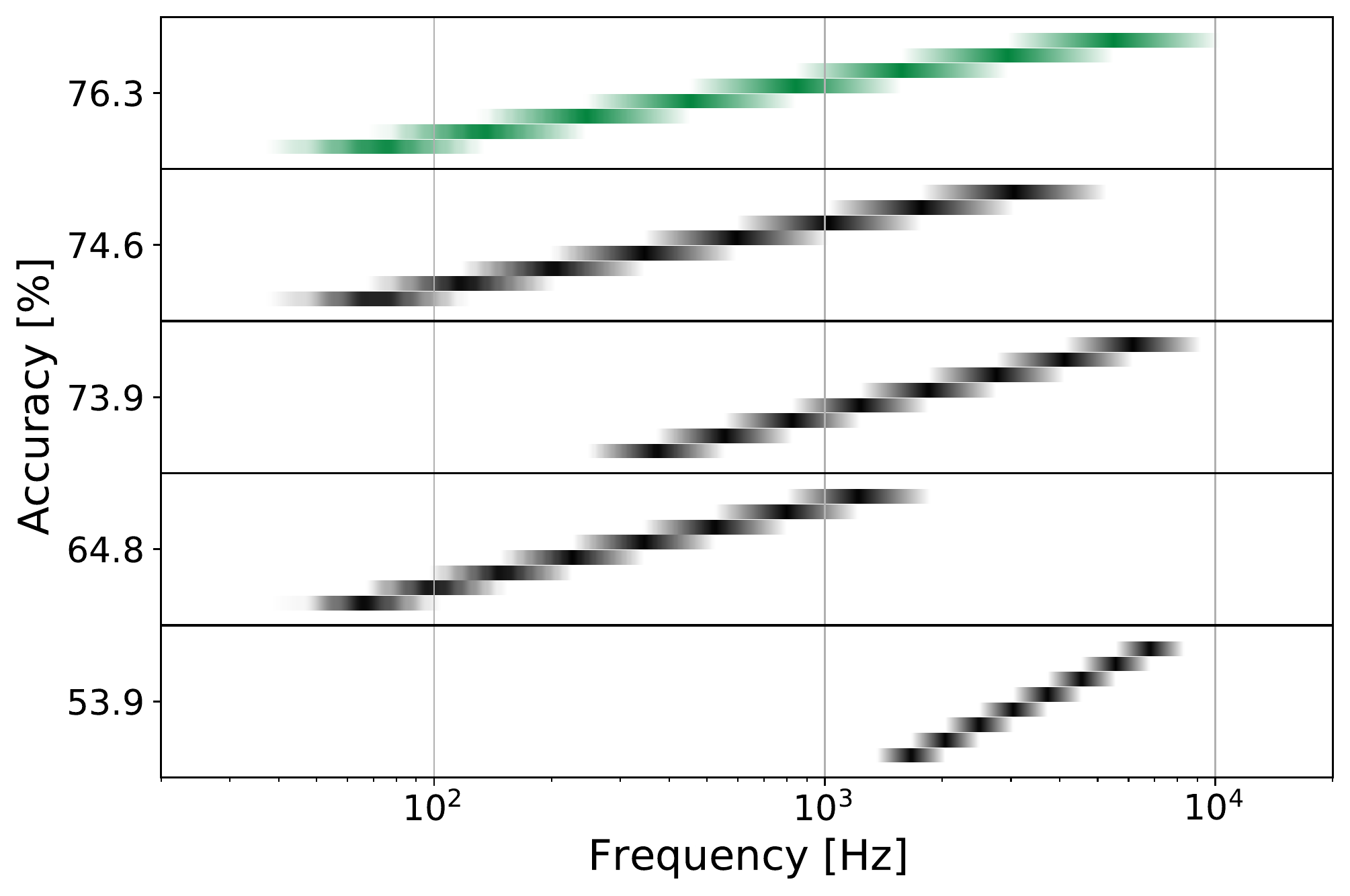}
\caption{Frequency range selection: each row corresponds to a model trained using the 8 filters indicated with the accuracy noted on the left. Their corner frequencies are equally spaced on the Mel scale. The highest-accuracy filter bank used throughout the paper is highlighted in green with corner frequencies 50\,Hz and 8\,kHz. A higher color intensity corresponds to lower attenuation of the signal at that frequency by the corresponding filter.}
\label{results:freqs}
\end{figure}

\figref{results:freqs} shows different accuracy of the different bars. Splitting the frequency range in low and high frequency and check which gives most of the information brings to another result: only the combination of high and low frequency gives the most relevant information for classification. The black bars confirm this trend: by enlarging the frequency range, the performances increase. Thus, the best performance is given by including the full frequency range.

\subsection{Accuracy-Energy Trade-Off}
\begin{figure}[tb]
  \includegraphics[width=1.0\columnwidth]{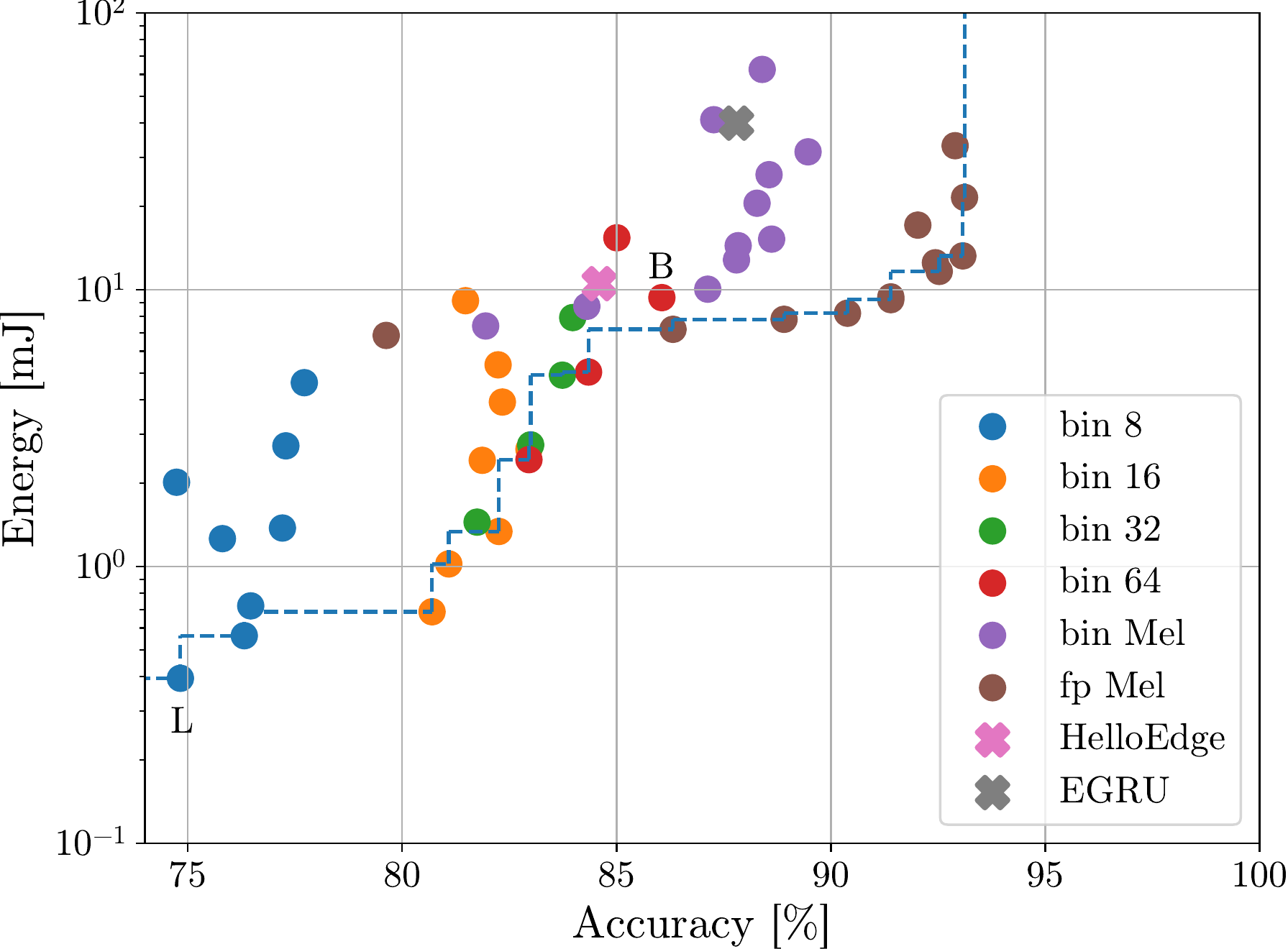}
  \caption{Pareto curve for different features. Multiple dots for a single feature represent a different neural network architecture, more specifically, different number of channels in the convolutional layers. The different colors indicate the HW implementation, particularly bin $k$, shows the implementation with $k$ analog filter banks. The best performing and lowest power AFE-based models are marked with B and L respectively. The frequency range of the filter is the one resulting the best from the analysis in \figref{results:freqs}, from 50 Hz to 8 kHz.} \label{fig:pareto}
\end{figure}

In this section, we present how the trade-off between power consumption and accuracy motivates the use of the analog front end. 
Therefore, we tried different analog front-end configurations, namely 8, 16, 32, and 64 filters. 
\figref{fig:pareto} shows the Pareto curve for the keyword spotting task. Here different configurations are presented, including MFCC and AFE features. 
The rightmost values are the network that uses full-precision numbers, and they perform the best accuracy with a power consumption of tens of mW. On their left, there are the systems that include the analog front end and binarization of weights ad activations. Here, the system achieves sub-mW consumption with 80\% accuracy in the task, 4\% less accuracy than the solution from HelloEdge \cite{zhang2017hello}, which consumes more than 10 mW.

The Pareto curve is divided into two clusters; below 85\% of accuracy, the use of the analog front end achieves better performance with respect to full-precision MFCC based features. The AFE does not reach a higher accuracy than 85\% because of loss of information in binarization. Therefore the Mel-based full-precision network achieves the best accuracy results.
More interesting is the violet cluster, where the audio is acquired in full-precision, and then Mel features are binarized. None of the points belong to the Pareto curve because the power consumption is too high due to the full-precision data acquisition and pre-processing, and full-precision networks with fewer parameters achieve the same accuracy performance with less energy.  
This insight demonstrates that BNN are particularly favourable with respect to full-precision networks when the acquisition process can be optimized when producing binary features. The analog front end is the enabling factor for the use of BNN in low-power contexts.

\subsection{Effect of Binarization on Different Classes}
\begin{figure*}[tb]
\includegraphics[width = \linewidth]{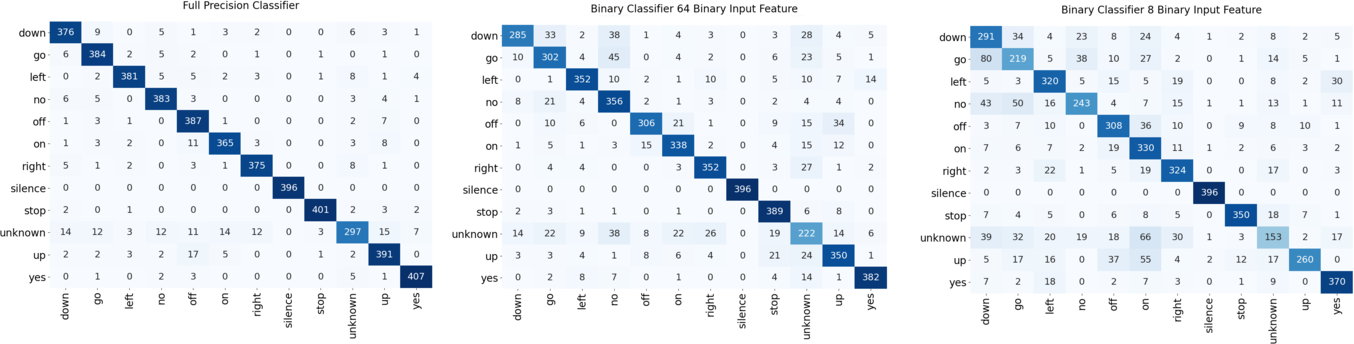}
  \caption{Confusion matrix of 1) the best-performing full-precision network, 2) the best performing solution with 64-filters AFE, and 3) the most energy efficient binary classifier with 8-filters AFE. The misclassifications in (2), (3) remain few and primarily affect the  most similar words (e.g., ``go'', ``down'', and ``no'') that could potentially be avoided in a user interface.} \label{fig:confusione}
\end{figure*}

To better understand how the AFE and binarization affect the various classes, \figref{fig:confusione} shows the confusion matrix for three different configurations. In the full precision system, the only class that shows different behavior to the other classes is the ``unknown'' class. It is the most challenging class because it contains 25 different words, among them some that are similar to the 10 target classes. 
As already shown in \figref{fig:pareto}, the BNN-based classifier with the binary AFE generally performs slightly worse in exchange for the clearly reduced energy cost. Interestingly, other than the ``unknown'' class, the errors are clustered among specific pairs of keywords, such as ``no'' and ``go'' or ``down'' and ``no''. This could be taken into consideration during the design of a user application, either by avoiding very similar keywords, or by providing context---certain circumstances might only permit a ``yes'' or ``no'' answer, or we might expect a keyword indicating a direction. 

\subsection{Comparison with Related Work}\label{sec:comparisonwithrelatedwork}
This section presents a comparison between our results and related work that uses the Google Speech Commands dataset to evaluate the keyword spotting system's performance. In order to make a direct comparison, we selected works that target microcontroller-based systems. 

\tblref{tbl:comparison} presents the breakdown of the energy consumption in terms of acquisition and processing. Among the models using the analog front end, we selected the classifiers with the lowest energy consumption and highest accuracy. As expected, the lowest energy consumption solution uses 8 AFE filters, while the highest accuracy uses 64 AFE filters. Our solution shows a substantial reduction in energy for the acquisition. At the same time, the best accuracy solution has a processing power in a similar range than Hello Edge. 

The total energy shows that our highest energy efficient solution has a power consumption of 1--2 orders of magnitude less, thus enabling scenarios in which the energy budget is below 1\,mJ. 
We compare the accuracy with both, Hello Edge and EGRU, where the first maintains the standard evaluation with the Google Speech Commands V2 dataset as described in \secref{sec:dataset} depicted in the dataset with 12 classes, and the second uses spoken digits, thus 10 classes.
We present numbers for both cases, showing that for the high-accuracy configuration we outperform the accuracy of either while maintaining a clearly reduced energy consumption. Even with the lowest-power configuration we can maintain a comparable accuracy to EGRU while dramatically reducing the power. 

We further show the importance of the low-power analog feature extraction, with the full-precision data acquisition and pre-processing alone taking 5.5\,mJ while the low-power configuration with our AFE uses only 0.56\,mJ in total including classification. 

\begin{table*}[tb]
\centering
\caption{State-of-the-Art Comparison with Fully-Digital Embedded Keyword Spotting Systems using 12 Keyword Classes and 10 Classes from the Google Speech Command Dataset}\label{tbl:comparison}
\begin{tabular}{lrrr@{\ }rr@{\ }rr@{\ }r} \toprule
\multirow{2}{*}{Method}                             & \multicolumn{4}{c}{----------------------- Energy [mJ] -----------------------} & \multicolumn{4}{c}{------------- Accuracy [\%] -------------} \\ 
                                   & Acquisition + pre-proc. & processing & \multicolumn{2}{c}{total (improv.)} & \multicolumn{2}{c}{12 classes} & \multicolumn{2}{c}{10 classes} \\ \midrule
Hello Edge \cite{zhang2017hello}                            & 5.48\,mJ                        & 5.04\,mJ                       & 10.52\,mJ & ($3.8\times$)                 & 84.6\% & (ref.)                & --  &                  \\
DS-CNN  \cite{reviewer2website}                                  & 5.48\,mJ                       & 5.57\,mJ                      & 11.05\,mJ & ($3.4\times$)                & 94.0\%  & ($+9.4\%$)                   & -- &                \\
EGRU  \cite{amoh2019optimized}                                  & 5.48\,mJ                        & 34.52\,mJ                      & 40.00\,mJ & (ref.)                & -- &                   & 87.8\% & (ref.)               \\
Our work (64-ch AFE, best acc.)            & 1.54\,mJ                        & 7.82\,mJ                       & \ 9.36\,mJ & ($4.3\times$)                  & 86.0\% & ($+1.4\%$)               & 88.8\% & ($+1.0\%$)     \\
Our work (8-ch AFE, lowest power) & 0.28\,mJ                        & 0.27\,mJ                       & \ 0.56\,mJ & ($71.4\times$)                  & 76.3\% & ($-8.3\%$)                & 85.8\% & ($-2.0\%$)               \\\bottomrule
\end{tabular}
\end{table*}

\section{Discussion and Comparison for VLSI Implementation}\label{sec:vlsidiscussion}
We evaluated the proposed method using an implementation without custom silicon. As summarized in \secref{sec:relWork_integrated}, fully-integrated implementations can massively reduce the power requirements. 
A direct comparison to these works is not meaningful, as essential contributors to the power consumption of a complete system are missing (power supply, clock generators, microphone, MCU \& peripherals to operate actors), and techniques such as extreme voltage scaling are incompatible with product-grade devices that need to operate reliably for several years. In order to enable a reasonable comparison, we estimate the expected power when integrating the proposed approach: 
\begin{enumerate}
    \item The AFE in \cite{fuketa2021time} is very similar---analog-domain band-pass filtering of audio signals and comparator-based conversion---and requires 0.14\,\si{\micro\watt} for 10 frequency bands. This includes some additional components that would not be used in our case. We can thus estimate 0.11\,\si{\micro\watt} for our 8-ch AFE and 0.89\,\si{\micro\watt} for our 64-ch AFE as an upper bound. 
    \item BNN inference is highly efficient and could be done with a specialized accelerator circuit such as CUTIE \cite{CUTIE} at 2\,POp/s/W for ternary NNs on average for a full network. For our BNN with 
    31.4\,MOp/classification, 
    this amounts to 15.7\,\si{\nano\joule}/classification.
\end{enumerate}
With no digital pre-processing required, we thus estimate the overall energy efficiency at around 0.92\,\si{\micro\watt} at a 2\,Hz classification rate for the 64-ch configuration with 86.0\% accuracy on 12 classes. Our approach can thus overcome the accuracy/task complexity limitations (88\% for 3 classes) of Fuketa \cite{fuketa2021time} while improving the energy efficiency over Vocell \cite{giraldo2020vocell} by $17\times$ (0.92\,\si{\micro\watt} v. 16\,\si{\micro\watt}) at an accuracy drop of 4.9\%, or compared to Giraldo et al.'s TCN accelerator \cite{giraldo2021efficient} $7.6\times$ (0.92\,\si{\micro\watt} v. 7\,\si{\micro\watt}) at an accuracy drop of 7.3\%. 

Our method could further be combined with Vocell's multi-stage approach to reduce power by skipping KWS classification when no sound is detected. This would allow to run our method with a more area-efficient BNN accelerator that is less energy efficient. Particularly in always-on applications where more than 90\% of the time no sound is detected, thus a circuit combining these approaches avoids the large permanent power contribution of the ADC and could lead to a highly efficient and tiny circuit with a sufficiently simple sound detector based on binary features. The full integration is left for future work.

\section{Conclusion}
We have explored the combination of analog binary feature extraction with a binarized neural network for energy-efficient keyword spotting on MCUs. This mix allows us to avoid that our energy cost is dominated by the data acquisition, leading to an efficiency gain of $4.6\times$ compared to previous work \cite{amoh2019optimized} while even slightly increasing the accuracy (+1\%). The elimination of most of the data acquisition energy allows us to further advance into the extremely low energy domain by trading off accuracy against energy as we adjust the BNN model complexity, i.e., a $71\times$ reduction in overall energy cost at a small accuracy drop of 2\%.

Previous approaches relying on the conventional ADC + preprocessing flow such as Hello Edge \cite{zhang2017hello} and EGRU, \cite{amoh2019optimized} even with an extremely optimized DNN model or hardware accelerator, will have an upper limit on the energy efficiency due to the data acquisition and pre-processing overhead. Our proposed system architecture with a binarizing AFE provides a method to overcome this fundamental limitation.

\section{Acknowledgments}
This work has been supported by the European Union’s Horizon 2020 research and innovation program under grant agreement no. 957337. This paper reflects only the authors' views and the European Commission cannot be held responsible for any use which may be made of the information contained therein.

\bibliographystyle{IEEEtran}
\bibliography{references, refs-additional}

\begin{thebibliography}{10}
\providecommand{\url}[1]{#1}
\csname url@samestyle\endcsname
\providecommand{\newblock}{\relax}
\providecommand{\bibinfo}[2]{#2}
\providecommand{\BIBentrySTDinterwordspacing}{\spaceskip=0pt\relax}
\providecommand{\BIBentryALTinterwordstretchfactor}{4}
\providecommand{\BIBentryALTinterwordspacing}{\spaceskip=\fontdimen2\font plus
\BIBentryALTinterwordstretchfactor\fontdimen3\font minus
  \fontdimen4\font\relax}
\providecommand{\BIBforeignlanguage}[2]{{%
\expandafter\ifx\csname l@#1\endcsname\relax
\typeout{** WARNING: IEEEtran.bst: No hyphenation pattern has been}%
\typeout{** loaded for the language `#1'. Using the pattern for}%
\typeout{** the default language instead.}%
\else
\language=\csname l@#1\endcsname
\fi
#2}}
\providecommand{\BIBdecl}{\relax}
\BIBdecl

\bibitem{michaely2017keyword}
A.~H. Michaely, X.~Zhang, G.~Simko, C.~Parada, and P.~Aleksic, ``Keyword
  spotting for google assistant using contextual speech recognition,'' in
  \emph{2017 IEEE Automatic Speech Recognition and Understanding Workshop
  (ASRU)}.\hskip 1em plus 0.5em minus 0.4em\relax IEEE, 2017, pp. 272--278.

\bibitem{mohammadi2018deep}
M.~Mohammadi, A.~Al-Fuqaha, S.~Sorour, and M.~Guizani, ``{Deep learning for IoT
  big data and streaming analytics: A survey},'' \emph{IEEE Communications
  Surveys {\textbackslash}{\&} Tutorials}, vol.~20, no.~4, pp. 2923--2960,
  2018.

\bibitem{takabi2010security}
H.~Takabi, J.~B.~D. Joshi, and G.-J. Ahn, ``{Security and privacy challenges in
  cloud computing environments},'' \emph{IEEE Security {\textbackslash}{\&}
  Privacy}, vol.~8, no.~6, pp. 24--31, 2010.

\bibitem{alioto2017iot}
M.~Alioto, ``{IoT: bird’s eye view, megatrends and perspectives},'' in
  \emph{Enabling the Internet of Things}.\hskip 1em plus 0.5em minus
  0.4em\relax Springer, 2017.

\bibitem{dillon2010cloud}
T.~Dillon, C.~Wu, and E.~Chang, ``{Cloud computing: issues and challenges},''
  in \emph{2010 24th IEEE international conference on advanced information
  networking and applications}.\hskip 1em plus 0.5em minus 0.4em\relax Ieee,
  2010, pp. 27--33.

\bibitem{yogi2017mist}
M.~K. Yogi, K.~Chandrasekhar, and G.~V. Kumar, ``{Mist computing: Principles,
  trends and future direction},'' \emph{arXiv preprint arXiv:1709.06927}, 2017.

\bibitem{premsankar2018edge}
G.~Premsankar, M.~Di~Francesco, and T.~Taleb, ``{Edge computing for the
  Internet of Things: A case study},'' \emph{IEEE Internet of Things Journal},
  vol.~5, no.~2, pp. 1275--1284, 2018.

\bibitem{krestinskaya2018binary}
O.~Krestinskaya and A.~P. James, ``{Binary weighted memristive analog deep
  neural network for near-sensor edge processing},'' in \emph{2018 IEEE 18th
  International Conference on Nanotechnology (IEEE-NANO)}.\hskip 1em plus 0.5em
  minus 0.4em\relax IEEE, 2018, pp. 1--4.

\bibitem{cerutti2020sound}
G.~Cerutti, R.~Andri, L.~Cavigelli, E.~Farella, M.~Magno, and L.~Benini,
  ``{Sound event detection with binary neural networks on tightly
  power-constrained IoT devices},'' in \emph{Proceedings of the ACM/IEEE
  International Symposium on Low Power Electronics and Design}, 2020, pp.
  19--24.

\bibitem{burr2021fair}
G.~W. Burr, S.~Lim, B.~Murmann, R.~Venkatesan, and M.~Verhelst, ``Fair and
  comprehensive benchmarking of machine learning processing chips,'' \emph{IEEE
  Design \& Test}, 2021.

\bibitem{panayotov2015librispeech}
V.~Panayotov, G.~Chen, D.~Povey, and S.~Khudanpur, ``{Librispeech: an asr
  corpus based on public domain audio books},'' in \emph{2015 IEEE
  international conference on acoustics, speech and signal processing
  (ICASSP)}.\hskip 1em plus 0.5em minus 0.4em\relax IEEE, 2015, pp. 5206--5210.

\bibitem{mesaros2018multi}
A.~Mesaros, T.~Heittola, and T.~Virtanen, ``{A multi-device dataset for urban
  acoustic scene classification},'' \emph{arXiv preprint arXiv:1807.09840},
  2018.

\bibitem{Sainat2015cnnkws}
T.~N. Sainath and C.~Parada, ``{Convolutional neural networks for
  small-footprint keyword spotting},'' in \emph{Proceedings of the Annual
  Conference of the International Speech Communication Association,
  INTERSPEECH}, vol. 2015-Janua, 2015, pp. 1478--1482.

\bibitem{lengerich2016end}
C.~Lengerich and A.~Hannun, ``{An end-to-end architecture for keyword spotting
  and voice activity detection},'' \emph{arXiv preprint arXiv:1611.09405},
  2016.

\bibitem{hwang2015online}
K.~Hwang, M.~Lee, and W.~Sung, ``{Online keyword spotting with a
  character-level recurrent neural network},'' \emph{arXiv preprint
  arXiv:1512.08903}, 2015.

\bibitem{pedroni2018small}
B.~U. Pedroni, S.~Sheik, H.~Mostafa, S.~Paul, C.~Augustine, and
  G.~Cauwenberghs, ``Small-footprint spiking neural networks for
  power-efficient keyword spotting,'' in \emph{2018 IEEE Biomedical Circuits
  and Systems Conference (BioCAS)}.\hskip 1em plus 0.5em minus 0.4em\relax
  IEEE, 2018, pp. 1--4.

\bibitem{Pandey2019}
A.~Pandey and D.~Wang, ``{TCNN: Temporal Convolutional Neural Network for
  Real-time Speech Enhancement in the Time Domain},'' in \emph{ICASSP 2019 -
  2019 IEEE International Conference on Acoustics, Speech and Signal Processing
  (ICASSP)}.\hskip 1em plus 0.5em minus 0.4em\relax IEEE, may 2019, pp.
  6875--6879.

\bibitem{Vaswani2017}
A.~Vaswani, N.~Shazeer, N.~Parmar, J.~Uszkoreit, L.~Jones, A.~N. Gomez,
  {\L}.~Kaiser, and I.~Polosukhin, ``{Attention is all you need},''
  \emph{Advances in Neural Information Processing Systems}, vol. 2017-Decem,
  no.~1, pp. 5999--6009, jun 2017.

\bibitem{tucker2016model}
G.~Tucker, M.~Wu, M.~Sun, S.~Panchapagesan, G.~Fu, and S.~Vitaladevuni,
  ``{Model Compression Applied to Small-Footprint Keyword Spotting.}'' in
  \emph{Interspeech}, 2016, pp. 1878--1882.

\bibitem{lin2016fixed}
D.~D. Lin, S.~S. Talathi, and V.~S. Annapureddy, ``{Fixed point quantization of
  deep convolutional networks},'' in \emph{33rd International Conference on
  Machine Learning, ICML 2016}, vol.~6, 2016, pp. 4166--4175.

\bibitem{Leng_2018}
C.~Leng, Z.~Dou, H.~Li, S.~Zhu, and R.~Jin, ``{Extremely low bit neural
  network: Squeeze the last bit out with ADMM},'' in \emph{32nd AAAI Conference
  on Artificial Intelligence, AAAI 2018}, 2018, pp. 3466--3473.

\bibitem{liu202022nm}
B.~Liu, H.~Cai, Z.~Wang, Y.~Sun, Z.~Shen, W.~Zhu, Y.~Li, Y.~Gong, W.~Ge,
  J.~Yang \emph{et~al.}, ``A 22nm, 10.8 $\mu$ w/15.1 $\mu$ w dual computing
  modes high power-performance-area efficiency domained background noise aware
  keyword-spotting processor,'' \emph{IEEE Transactions on Circuits and Systems
  I: Regular Papers}, vol.~67, no.~12, pp. 4733--4746, 2020.

\bibitem{AojunZhou2016}
A.~Zhou, A.~Yao, Y.~Guo, L.~Xu, and Y.~Chen, ``{Incremental Network
  Quantization: Towards Lossless CNNs with Low-Precision Weights},'' in
  \emph{Proc. ICLR}, 2017.

\bibitem{zhang2017hello}
Y.~Zhang, N.~Suda, L.~Lai, and V.~Chandra, ``{Hello Edge: Keyword Spotting on
  Microcontrollers},'' \emph{arXiv preprint arXiv:1711.07128}, 2017.

\bibitem{Courbariaux2016}
M.~Courbariaux and {others}, ``{Binarized Neural Networks: Training Deep Neural
  Networks with Weights and Activations Constrained to +1 or -1},'' in
  \emph{arXiv:1602.02830}, 2016.

\bibitem{rastegari2016xnor}
M.~Rastegari, V.~Ordonez, J.~Redmon, and A.~Farhadi, ``{Xnor-net: Imagenet
  classification using binary convolutional neural networks},'' in
  \emph{European Conference on Computer Vision}.\hskip 1em plus 0.5em minus
  0.4em\relax Springer, 2016, pp. 525--542.

\bibitem{hubara2016binarized}
I.~Hubara and {others}, ``{Binarized neural networks},'' in \emph{Adv. NIPS},
  2016, pp. 4107--4115.

\bibitem{zhou2016dorefa}
S.~Zhou and {others}, ``{Dorefa-net: Training low bitwidth convolutional neural
  networks with low bitwidth gradients},'' \emph{arXiv:1606.06160}, 2016.

\bibitem{spallanzani2019additive}
M.~Spallanzani and {others}, ``{Additive noise annealing and approximation
  properties of quantized neural networks},'' \emph{arXiv:1905.10452}, 2019.

\bibitem{gruenstein2017cascade}
A.~Gruenstein, R.~Alvarez, C.~Thornton, and M.~Ghodrat, ``{A cascade
  architecture for keyword spotting on mobile devices},'' \emph{arXiv preprint
  arXiv:1712.03603}, 2017.

\bibitem{amoh2019optimized}
J.~Amoh and K.~M. Odame, ``{An optimized recurrent unit for ultra-low-power
  keyword spotting},'' \emph{Proceedings of the ACM on Interactive, Mobile,
  Wearable and Ubiquitous Technologies}, vol.~3, no.~2, pp. 1--17, 2019.

\bibitem{meyer2017efficient}
\BIBentryALTinterwordspacing
M.~Meyer, L.~Cavigelli, and L.~Thiele, ``{Efficient Convolutional Neural
  Network For Audio Event Detection},'' \emph{arXiv preprint arXiv:1709.09888},
  2017. [Online]. Available: \url{http://arxiv.org/abs/1709.09888}
\BIBentrySTDinterwordspacing

\bibitem{cerutti2020compact}
G.~Cerutti, R.~Prasad, A.~Brutti, and E.~Farella, ``{Compact recurrent neural
  networks for acoustic event detection on low-energy low-complexity
  platforms},'' \emph{IEEE Journal of Selected Topics in Signal Processing},
  2020.

\bibitem{wen2006360}
B.~Wen and K.~Boahen, ``{A 360-channel speech preprocessor that emulates the
  cochlear amplifier},'' in \emph{2006 IEEE International Solid State Circuits
  Conference-Digest of Technical Papers}.\hskip 1em plus 0.5em minus
  0.4em\relax IEEE, 2006, pp. 2268--2277.

\bibitem{fragniere2005100}
E.~Fragni{\`{e}}re, ``{A 100-channel analog CMOS auditory filter bank for
  speech recognition},'' in \emph{ISSCC. 2005 IEEE International Digest of
  Technical Papers. Solid-State Circuits Conference, 2005.}\hskip 1em plus
  0.5em minus 0.4em\relax IEEE, 2005, pp. 140--589.

\bibitem{lyon1988analog}
R.~F. Lyon and C.~Mead, ``{An analog electronic cochlea},'' \emph{IEEE
  Transactions on Acoustics, Speech, and Signal Processing}, vol.~36, no.~7,
  pp. 1119--1134, 1988.

\bibitem{6869048}
S.~Wang, T.~J. Koickal, A.~Hamilton, R.~Cheung, and L.~S. Smith, ``{A
  Bio-Realistic Analog CMOS Cochlea Filter With High Tunability and Ultra-Steep
  Roll-Off},'' \emph{IEEE Transactions on Biomedical Circuits and Systems},
  vol.~9, no.~3, pp. 297--311, 2015.

\bibitem{5537160}
T.~Delbruck, T.~Koch, R.~Berner, and H.~Hermansky, ``{Fully integrated 500uW
  speech detection wake-up circuit},'' in \emph{Proceedings of 2010 IEEE
  International Symposium on Circuits and Systems}, 2010, pp. 2015--2018.

\bibitem{li2021ns}
Q.~Li, C.~Liu, P.~Dong, Y.~Zhang, T.~Li, S.~Lin, M.~Yang, F.~Qiao, Y.~Wang,
  L.~Luo \emph{et~al.}, ``Ns-fdn: Near-sensor processing architecture of
  feature-configurable distributed network for beyond-real-time always-on
  keyword spotting,'' \emph{IEEE Transactions on Circuits and Systems I:
  Regular Papers}, 2021.

\bibitem{yang20181muw}
M.~Yang, C.-H. Yeh, Y.~Zhou, J.~P. Cerqueira, A.~A. Lazar, and M.~Seok, ``{A
  1{\$}{$\mu$}{\$}W voice activity detector using analog feature extraction and
  digital deep neural network},'' in \emph{2018 IEEE International Solid-State
  Circuits Conference-(ISSCC)}.\hskip 1em plus 0.5em minus 0.4em\relax IEEE,
  2018, pp. 346--348.

\bibitem{giraldo2020vocell}
J.~S.~P. Giraldo, S.~Lauwereins, K.~Badami, and M.~Verhelst, ``Vocell: A 65-nm
  speech-triggered wake-up soc for 10-$\mu$ w keyword spotting and speaker
  verification,'' \emph{IEEE Journal of Solid-State Circuits}, vol.~55, no.~4,
  pp. 868--878, 2020.

\bibitem{giraldo2021efficient}
J.~Giraldo, V.~Jain, and M.~Verhelst, ``Efficient execution of temporal
  convolutional networks for embedded keyword spotting,'' \emph{IEEE
  Transactions on Very Large Scale Integration (VLSI) Systems}, 2021.

\bibitem{shan202014}
W.~Shan, M.~Yang, J.~Xu, Y.~Lu, S.~Zhang, T.~Wang, J.~Yang, L.~Shi, and
  M.~Seok, ``14.1 a 510nw 0.41 v low-memory low-computation keyword-spotting
  chip using serial fft-based mfcc and binarized depthwise separable
  convolutional neural network in 28nm cmos,'' in \emph{2020 IEEE International
  Solid-State Circuits Conference-(ISSCC)}.\hskip 1em plus 0.5em minus
  0.4em\relax IEEE, 2020, pp. 230--232.

\bibitem{chong20212}
Y.~S. Chong, W.~L. Goh, V.~P. Nambiar, and A.~T. Do, ``A 2.5 $\mu$w kws engine
  with pruned lstm and embedded mfcc for iot applications,'' \emph{IEEE
  Transactions on Circuits and Systems II: Express Briefs}, 2021.

\bibitem{fuketa2021time}
H.~Fuketa, ``Time-delay-neural-network-based audio feature extractor for
  ultra-low power keyword spotting,'' \emph{IEEE Transactions on Circuits and
  Systems II: Express Briefs}, 2021.

\bibitem{9429864}
M.~Yang, H.~Liu, W.~Shan, J.~Zhang, I.~Kiselev, S.~J. Kim, C.~Enz, and M.~Seok,
  ``Nanowatt acoustic inference sensing exploiting nonlinear analog feature
  extraction,'' \emph{IEEE Journal of Solid-State Circuits}, vol.~56, no.~10,
  pp. 3123--3133, 2021.

\bibitem{mayer2019self}
P.~Mayer, M.~Magno, and L.~Benini, ``{Self-sustaining acoustic sensor with
  programmable pattern recognition for underwater monitoring},'' \emph{IEEE
  Transactions on Instrumentation and Measurement}, vol.~68, no.~7, pp.
  2346--2355, 2019.

\bibitem{warden2018speech}
P.~Warden, ``{Speech commands: A dataset for limited-vocabulary speech
  recognition},'' \emph{arXiv preprint arXiv:1804.03209}, 2018.

\bibitem{GapNews2018}
{VentureBeat.com}, ``{GreenWaves Technologies unveils Gap8 processor for AI at
  the edge},'' 2018.

\bibitem{hershey2017cnn}
S.~Hershey, S.~Chaudhuri, D.~P. Ellis, J.~F. Gemmeke, A.~Jansen, R.~C. Moore,
  M.~Plakal, D.~Platt, R.~A. Saurous, B.~Seybold, M.~Slaney, R.~J. Weiss, and
  K.~Wilson, ``{CNN architectures for large-scale audio classification},'' in
  \emph{ICASSP, IEEE}.\hskip 1em plus 0.5em minus 0.4em\relax IEEE, 2017, pp.
  131--135.

\bibitem{reviewer2website}
``The speed and power advantage of a purpose-built neural compute engine,''
  \url{https://www.syntiant.com/post/keyword-spotting-power-comparison},
  accessed: 2021-12-10.

\bibitem{CUTIE}
M.~Scherer, G.~Rutishauser, L.~Cavigelli, and L.~Benini, ``Cutie: Beyond
  petaop/s/w ternary dnn inference acceleration with better-than-binary energy
  efficiency,'' \emph{IEEE Transactions on Computer-Aided Design of Integrated
  Circuits and Systems}, pp. 1--1, 2021.

\end{thebibliography}

\begin{IEEEbiography}[{\includegraphics[width=1in,height=1.25in,clip,keepaspectratio]{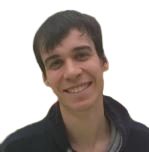}}]{Gianmarco Cerutti}
Recieved the B.Sc., M.Sc. degree in Electronic Engineering at the University La Sapienza in 2015 and 2017 respectively. In 2020 he has concluded his Ph.D in the Fondazione Bruno Kessler in Trento, Italy and at the Department of Electrical, Electronic and Information Engineering  Guglielmo Marconi, University of Bologna, under the supervision of Dr. Elisabetta Farella. His current research in Fondazione Bruno Kessler focuses on audio classification on resource constrained devices, embedded systems and artificial intelligence.
\end{IEEEbiography}
\begin{IEEEbiography}[{\includegraphics[width=1in,height=1.25in,clip,keepaspectratio]{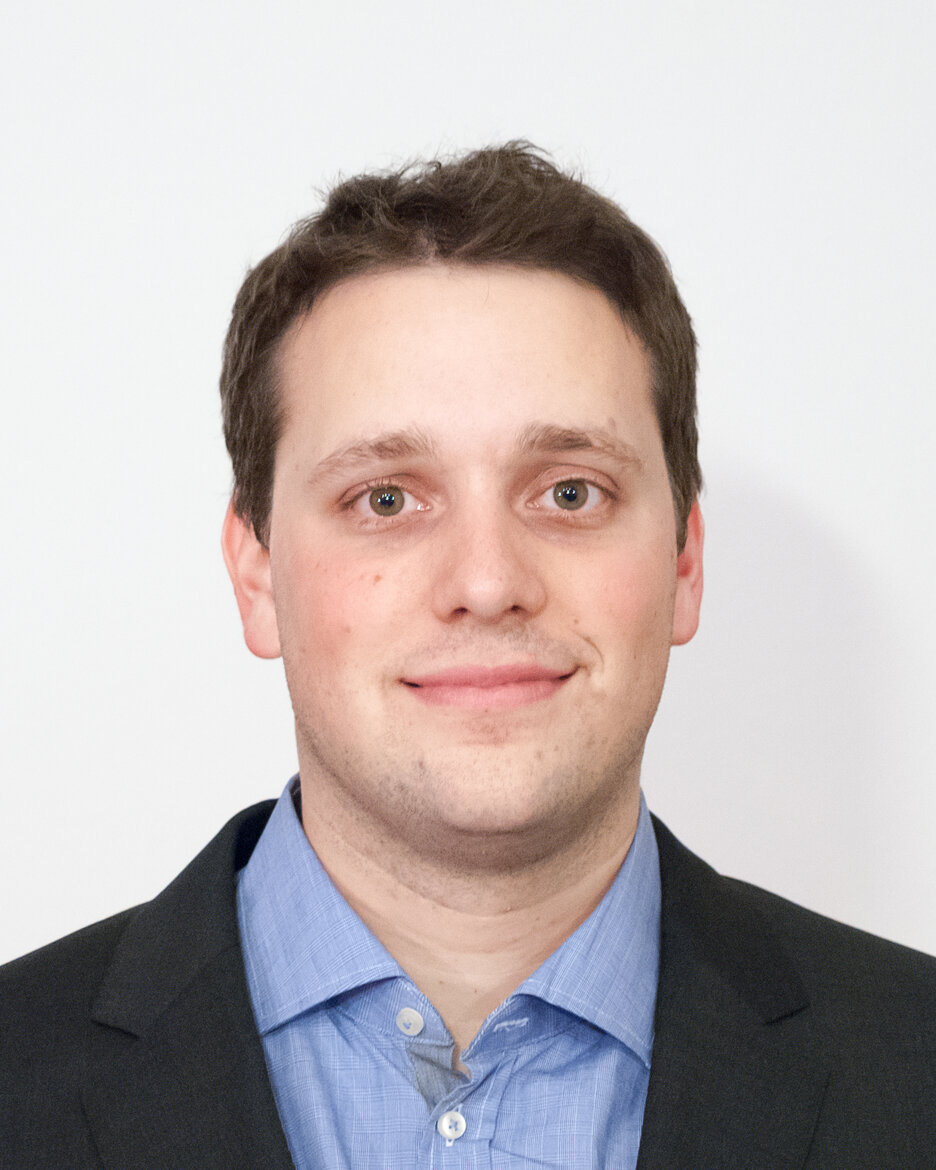}}]{Lukas Cavigelli}
received the B.Sc., M.Sc., and Ph.D. degree in electrical engineering and information technology from ETH Zurich, Switzerland in 2012, 2014 and 2019, respectively. After spending a year as a Postdoc at ETH Zurich, he has joined Huawei's Zurich Research Center in Spring 2020. His research interests include deep learning, computer vision, embedded systems, and low-power integrated circuit design. He has received the best paper award at the VLSI-SoC and the ICDSC conferences in 2013 and 2017, the best student paper award at the Security+Defense conference in 2016, and the Donald O. Pederson best paper award (IEEE TCAD) in 2019.
\end{IEEEbiography}
\begin{IEEEbiography}[{\includegraphics[width=1in,height=1.25in,clip,keepaspectratio]{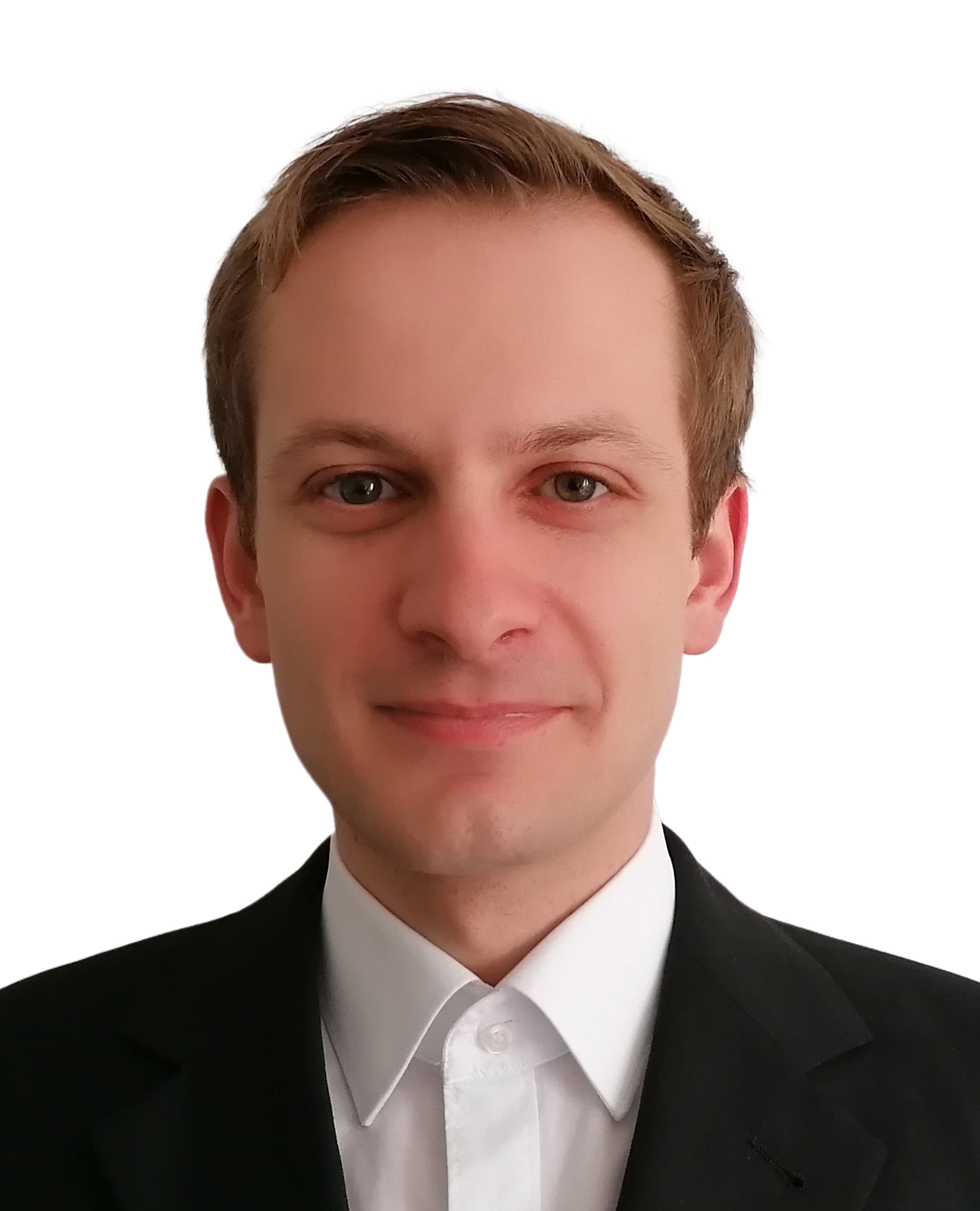}}]{Renzo Andri}
received the B.Sc., M.Sc. and Ph.D. degree in Electrical Engineering and Information Technology at ETH Zurich in 2013, 2015, and 2020, respectively. His research focuses on energy-efficient machine learning acceleration from embedded system design to full-custom IC design. He is currently working as a senior researcher at Huawei Research Center Zurich. In 2019, he has won the IEEE TCAD Donald O. Pederson Award.
\end{IEEEbiography}
\begin{IEEEbiography}[{\includegraphics[width=1in,height=1.25in,clip,keepaspectratio]{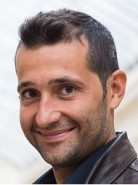}}]{Michele Magno}
is currently a Senior Researcher and Lecturer at ETH Zürich, Switzerland, at the Department of Information Technology and Electrical Engineering (D-ITET). Since 2020 is leading the D-ITET center for project-based learning. He received his master and Ph.D. degrees in electronic engineering from the University of Bologna, Italy, in 2004 and 2010, respectively. He is working in ETH since 2013 and has become a visiting lecturer or professor in several universities, namely the University of Nice Sophia, France, Enssat Lannion, France, University of Bologna and Mid University Sweden.
His current research interests include smart sensing, low power machine learning, wireless sensor networks, wearable devices, energy harvesting, low power management techniques, and extension of the lifetime of batteries-operating devices. He has authored more than 150 papers in international journals and conferences. Some of his publications were awarded as best papers awards in IEEE conferences such as IEEE International Conference on E-health Networking, Application \& Services 2018, IEEE Sensors Applications Symposium (SAS) 2018, IEEE International Workshop on Advances in Sensors and Interfaces 2017 among others. 
\end{IEEEbiography}
\begin{IEEEbiography}[{\includegraphics[width=1in,height=1.25in,clip,keepaspectratio]{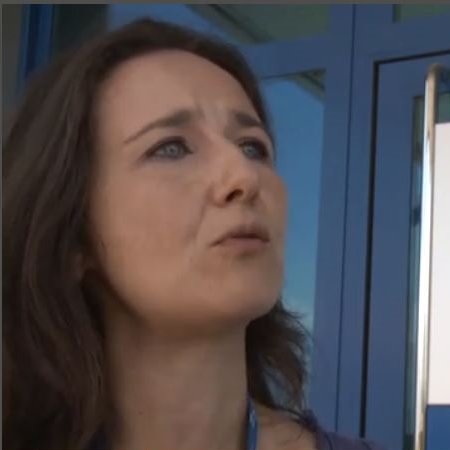}}]{Elisabetta Farella}
Ph.D., tenured researcher and head of E3DA (Energy Efficient Embedded Digital Architecture) research unit (e3da.fbk.eu) at FBK (Fondazione Bruno Kessler) from 2014. She was previously a research fellow at DEI, University of Bologna, where she has been an adjunct Professor. She was also research fellow at CINECA Supercomputing Center and research supervisor at T3LAB, a technology transfer initiative promoted by Confindustria Bologna. Her research is in the field of energy-independent embedded systems that are, at the same time, equipped with artificial on-board intelligence. Examples of such systems are wireless sensor networks, wearable electronics, Internet of Things from the point of view of energy efficient devices equipped with smart sensors and actuators. These technologies are used in various application fields from motor rehabilitation to human-machine interaction, in smart cities and communities, etc. She is involved in a number of international conferences and networks and co-author of about 150 papers in major International conferences and journals in the field of embedded systems. 
\end{IEEEbiography}
\begin{IEEEbiography}[{\includegraphics[width=1in,height=1.25in,clip,keepaspectratio]{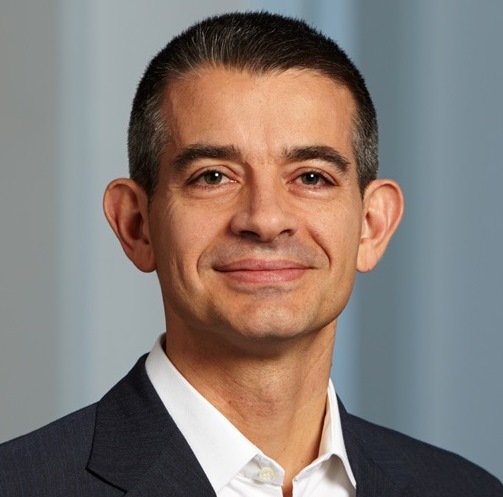}}]{Luca Benini}
is the Chair of Digital Circuits and Systems at ETH Zürich and a Full Professor at the University of Bologna. He has served as Chief Architect for the Platform2012 in STMicroelectronics, Grenoble. Dr. Benini’s research interests are in energy-efficient system and multi-core SoC design. He is also active in the area of energy-efficient smart sensors and sensor networks. He has published more than 1’000 papers in peer-reviewed international journals and conferences, four books and several book chapters. He is a Fellow of the ACM and of the IEEE and a member of the Academia Europaea.
\end{IEEEbiography}

\end{document}